\documentclass[12pt]{article}
\usepackage{graphicx}
\usepackage{amsmath}
\usepackage{amsfonts}
\usepackage{amssymb}

\usepackage{authblk}

\author[1]{Ryosuke Ishii}
\author[2]{Kuninori Nakagawa\footnote{Corresponding author, email to nakagawa.kuninori@shizuoka.ac.jp}}
\affil[1]{Teikyo University, Otsuka 359, Hachioji, Tokyo 1920395, Japan.}
\affil[2]{Shizuoka University, Ohya 836, Suruga-ku, Shizuoka 4228529, Japan.}
\begin{document}

\title{Government Expenditure on Research Plans and their Diversity\footnote{The authors are grateful to Keisuke Kawachi and Noriaki Matsushima for useful discussion and suggestion of the topic treated in this paper at ARSC2017 at the University of Tokyo. We would also like to thank Hiroshi Aiura and seminar participants at Nanzan University for helpful comments.}}
\date{\today}  
\maketitle

\begin{abstract}
In this study, we consider research and development investment by the government. Our study is motivated by the bias in the budget allocation owing to the competitive funding system. In our model, each researcher presents research plans and expenses, and the government selects a research plan in two periods---before and after the government knows its favorite plan---and spends funds on the adopted program in each period. We demonstrate that, in a subgame perfect equilibrium, the government adopts equally as many active plans as possible. In an equilibrium, the selected plans are distributed proportionally. Thus, the investment in research projects is symmetric and unbiased. Our results imply that equally widespread expenditure across all research fields is better than the selection of and concentration in some specific fields.
\end{abstract}

\section{Introduction}
In Japan, since 1995, research and development (R$\&$D) investment by the government has been promoted by the Science and Technology Basic Plan, which is a five-year plan of the Japanese government. Currently, the plan has entered its fifth phase, with its latest version issued in 2016. At the same time, the government's expenditure is bound by the Act on Special Measures Concerning Promotion of Fiscal Structural Reform (Act No. 109 of 1997). In particular, subject to the provisions of Article 25 Paragraph 2 of the Act, the government should promote the selection of R$\&$D investment and improve its efficiency by appropriately evaluating the results and reflecting them in its budget allocation. That is, the government focuses on the selection of and concentration in some specific fields. Although the Japanese government has recognized the role of government investment in ensuring long-term growth of the economy, it adopts policies of selection and concentration owing to financial problems. Policymakers want to have the ability to identify a winning lottery in advance. We consider the best policy for the government in this situation.

We consider R$\&$D investment by the government. Theoretically, we extend the work of Ishii and Nakagawa (2015)\cite{ishiinakagawa}. In our model, the government is an investor to whom researchers pitch their plans. Each researcher presents his or her research plans along with the corresponding expenses, and the government selects a research plan in each of two periods---before and after the government knows its favorite plan---and spends funds on the adopted program in each period. The optimal research outcome for the government is not known ``ex ante" before the government knows its favorite plan, while it is known ``ex post" after the government knows its favorite plan. In other words, the government does not know the number of winning lots in advance but learns it in the aftermath of a situation. 
Here, we consider national expenditure on research to denote government's ``purchase" of a research plan, as the cost of research for the government is the ``price" of research for the researchers. In addition, our model distinguishes between ``purchases" and ``consumption." Although the government spends funds on various types of academic research, except for ideal research, the government later discards it, regarding it as useless for the nation after learning its favorite plan. In other words, purchase of unconsumed research by the government results only in incurred costs as this research is useless and its consumption does not generate any positive utility. 

We analyze the variety of research that is determined endogenously and the degree of differentiation in research projects across the varieties. We demonstrate that, in a subgame perfect equilibrium, the investment in research projects is symmetric and unbiased. In other words, the government's expenditure is spread out over all scientific fields. In a subgame perfect equilibrium, the selected plans are distributed proportionally. It is optimal for the government to purchase as many active plans as possible. Our results imply that equally widespread expenditure across all research fields is better than the selection of and concentration in some specific fields. In addition, we calculate the optimal variety at equilibrium, which is the optimal number of research projects for the government. 

\section{Model}

Each researcher $i$ chooses once simultaneously his or her own ``location" of characteristic, and then chooses twice simultaneously a ``price" for the plan. Researchers set the characteristics and prices of their research plans ex ante and then set their prices again ex post. Between ex ante and ex post, $t$, the ideal point of the government is realized. $Z=(z_{1},z_{2},\ldots,z_{n})$ denotes the characteristic vector of the researchers' plans. A researcher $i(=1,2,\ldots,n)$ chooses $z_{i}$ in the interval $[0,1]$, which is his or her plan's characteristic. Let $p_i$ denote the price of the researcher's plan with characteristic $z_{i}$. Here, we define $p_{i}:Z\rightarrow \lbrack0,\infty)$. 

The government selects a research plan in two periods---before, $b$, and after, $a$, the government knows its favorite plan---and spends funds on the adopted program in each period. The government chooses a subset $I\subseteq M$, where $M$ denotes the whole set of the research plan $M=\{1,2,\cdots,n\}$. We define the government's utility $u$ as $\bar{u}-(t-z_{i})^{2}-($total expenditure for adopted plans$)$ if researcher $i$'s plan is adopted. $E[(t-z_{i})^{2}]$ is the government's expected utility loss when it adopts plans before its ideal points are known. We assume that $\bar{u}$ is large enough for the government to adopt at least one research plan. This utility function is consistent with Ishii and Nakagawa (2015)\cite{ishiinakagawa}. 
 
Each researcher obtains a fund when his or her plan is accepted. Researcher $i$'s gain $\pi_{i}$ is
\[
\pi_{i}=\left\{
\begin{array}
[c]{l}%
p_{i}^b\quad\text{if they are adopted ex ante,}\\
p_{i}^a\quad\text{if they are adopted ex post,}\\
0\quad\text{if they are not adopted.}%
\end{array}
\right.
\]

\section{Ex-Post Price Game}

After the government's ideal point $t$ is realized, we can classify its realization in the interval $[0,1]$ into the following three cases: (1) $0 \le t \le z_{1}$, (2) $z_{i} \le t \le z_{i+1}, (1 \le i \le n-1)$, and (3) $z_{n} \le t \le1$. Here,  without loss of generality, we focus on the case in which $t$ is realized in $z_{i} \le t \le (z_{i}+z_{i+1})/2 \le z_{i+1}$. In other words, for simplicity, we assume that $z_{i}$ is the closest point from the realized $t$. In this case, the government's most preferred plan is $z_{i}$. Thus, the government will not need any more plans ex post if they already have a plan, $z_{i}$, before they find their ideal point, $t$.

Next, we consider the case in which the government does not have the most favorite plan, $z_{i}$, ex ante. In this scenario, we have to consider the following two cases: one in which the government has already chosen one or more plans but has not chosen plan $z_{i}$ ex ante, and the other in which the government chooses no plan ex ante. In addition, the former is divided into two scenarios: first, the government already has $z_{i+1}$ and, second, it does not have $z_{i+1}$. 

First, the government does not have plan $z_{i}$ ex ante, that is, it has not an ideal plan. Suppose that among the government's adopted plans in advance, in terms of characteristics, plan $z_j \ne z_i$ is the closest to the government's realized ideal point $t$. Now, we show that the unique equilibrium outcome in this case is that the government adopts an additional plan $z_{i}$ ex post. To that end, we have to demonstrate that the government does not choose any plans except for $z_{i}$ in the ex-post price subgame, that is, no plan except $z_{i}$ is demanded ex post.　

At equilibrium, the government purchases its most preferred plan $z_{i}$ at a price equal to the margin of transportation cost in both cases wherein $z_j = z_{i+1}$ and $z_j \neq z_{i+1}$. Roughly speaking, the government discards the losing lot and repurchases the winning lottery. Now, we assume that the government already has plan $z_j$ ex ante when $z_i$ is closest to its ideal point, which is realized ex post. In addition, we assume that the government already has $z_j = z_{i+1}$ for the time being.

Using a proof by contradiction, suppose that the government does not adopt an additional plan $z_{i}$ at the equilibrium, and then it chooses plan $z_j \ne z_i$, which is second-closest to the government's realized ideal point $t$ in terms of the characteristics at that time. The government's utility is 
\begin{equation}
u-(t-z_{j})^{2}-\text{(total expenditure ex ante)}.%
\end{equation}
On the contrary, if the government adopts plan $z_i$ with $p_i^a$ ex post, its utility is
\begin{equation}
u-(t-z_{i})^{2}-p_{i}^{a}-\text{(total expenditure ex ante)}.%
\end{equation}
When $p_{i}^{a}$ satisfies the following condition
\begin{equation}
p_{i}^{a}<(t-z_{j})^{2}-(t-z_{i})^{2}, \label{expost}%
\end{equation}
the government's ex-post utility is strictly improved. On the contrary, researcher $i$'s profit is strictly improved if the government chooses it with a positive $p_{i}^{a}$ when its price changes such that \eqref{expost} holds. This is in contradiction to the fact that the government does not adopt an additional plan $z_{i}$ at the equilibrium. In other words, the government discards the losing lot $z_j$ and repurchases the winning lottery $z_i$ at the equilibrium. 

Next, we consider the case in which the government already has plan $j$ with characteristic $z_j \neq z_{i+1}$, for example, $z_j = z_{i-1}$. In this case, compared to $z_j=z_{i+1}$, every $z_j \neq z_{i+1}$ is strictly far away from $z_i$. Now, we check that, at an ex-post equilibrium, the government does not choose $z_k$, which meets the following condition: $(t-z_i)^2 < (t-z_k)^2 < (t-z_j)^2$. Suppose that the government additionally purchases plan $k \neq i$ at an ex-post equilibrium. It obtains $u-(t-z_{k})^{2}-p_{k}^{a}-\text{(total expenditure at ex ante)}$. On the other hand, when the government additionally purchases plan $i$, it obtains $u-(t-z_{i})^{2}-p_{i}^{a}-\text{(total expenditure at ex ante)}$. Now, at this ex-post equilibrium, the following holds:
\begin{equation}
p_{i}^{a}\geq(t-z_{k})^{2}-(t-z_{i})^{2}+p_{k}^{a},%
\end{equation}
This is because the government chooses $k$ and not $i$. However, $i$ has the incentive to change its price from $p_i^{a}$ to $p_i^{a\prime}$ such that
\begin{equation}
p_{i}^{a\prime}<(t-z_{k})^{2}-(t-z_{i})^{2}+p_{k}^{a} \label{expost2}.%
\end{equation}
Thus, the government additionally purchases the best-fitted plan $i$ at an ex-post equilibrium. 

Thus, we find that, after the government finds that the characteristic of the plan possessed by the government is not best fitted to the realized ideal point $t$, the government purchases an additional plan that is closest to its realized ideal point $t$. Moreover, we also find that, when the government has no plan in advance, it chooses a plan that is the closest to its realized ideal point $t$ after it learns its ideal point.

Finally, we calculate the ex-post equilibrium price $p_{i}^{a\ast}$. As mentioned previously, we have to consider the two cases in which the government already has plan $z_j = z_{i+1}$ or $z_j \neq z_{i+1}$ when it has plan $j$ with $z_{j}$. We obtain the ex-post equilibrium price as follows: $p_{i}^{a\ast}=(t-z_{i+1})^{2}-(t-z_{i})^{2}$ and $p_{i+1}^{a\ast}=0$. 

It is the best response for researcher $i$ to charge an ex-post price $p_{i}^{a}$ such that $p_{i}^{a}\leq(t-z_{i+1})^{2}-(t-z_{i})^{2}+p_{i+1}^{a}$ holds. In other words, researcher $i$ sets its price as high as the government can afford. In particular, researcher $i$ charges a price $p_{i}^{a}$ that is equal to the price at which the government is indifferent between plan $z_i$ and plan $z_{i+1}$, whose characteristic is closest to plan $z_i$. Thus, we obtain $p_{i}^{a\ast}=(t-z_{i+1})^{2}-(t-z_{i})^{2}+p_{i+1}^{a}$.

Next, we find that $p_{i+1}^{a\ast}=0$. Suppose that $p_{i+1}^{a}>0$. This presents a contradiction. This is because researcher $i+1$ can charge price $p_{i+1}^{a}-\varepsilon>0$ to obtain more profits, thus making the government choose the plan. Thus, we obtain $p_{i+1}^{a\ast}=0$. This yields $p_{i}^{a\ast}=(t-z_{i+1})^{2}-(t-z_{i})^{2}$. Furthermore, given the price pair, when the government chooses $i$'s plan, no plan deviates from this price pair. Thus, we find that this pair is the equilibrium price pair ex post.

The previous discussion with regard to $z_j =  z_{i+1}$ that is presented in this section applies to the case of $z_j \neq z_{i+1}$. We can set $p_{j}^{a\ast}=0$ that is second-closest to the government's realized ideal point $t$ in terms of characteristics. 
Thus, we obtain
\begin{equation}\label{exposteq}
p_{i}^{a\ast}=(t-z_{i+1})^{2}-(t-z_{i})^{2}. 
\end{equation}

\section{Ex-Ante Price Game}

In particular, on a realized path of equilibrium, all researchers set the ex-ante prices of their research plans so that they obtain profits that equal the expected profit they would obtain ex post and the government chooses all the plans or a part of them. In this section, we demonstrate that this is the behavior at the equilibrium of an ex-ante price subgame, where the researchers obtain the same expected equilibrium profit.

Our analysis proceeds as follows. First, we demonstrate that it is not optimal for the government to purchase a plan that gives a researcher a higher profit than his or her expected profit ex post. Next, we demonstrate that when a researcher charges a price lower than his or her expected profit ex post, a unique best response for the government is to purchase the researcher's plans. In other words, the government does not choose a plan that is relatively high in price, but chooses a plan with a relatively low price. Finally, none of the researchers set their prices lower than their expected profit obtained in the ex-post price game. In summary, we demonstrate that the government chooses all the plans or a part of them for which researchers set the prices ex ante such that they obtain a profit equal to the expected profit ex post.  

Using the equilibrium price in the ex-post price subgame, we obtain the ex-post equilibrium profit as follows: 
\begin{align}\label{9}
\pi_i^{a\ast} = p_{i}^{a\ast} 
&  =\max\{0,\min\{(t-z_{i-1})^{2},(t-z_{i+1})^{2}\}-(t-z_{i})^{2}\}\\
&  =\left\{
\begin{tabular}
[c]{lll}%
$\left(  t-z_{i-1}\right)  ^{2}-\left(  t-z_{i}\right)  ^{2}$ & if &
$\frac{z_{i-1}+z_{i}}{2}\leq t\leq\frac{z_{i-1}+z_{i+1}}{2},$\\
$\left(  t-z_{i+1}\right)  ^{2}-\left(  t-z_{i}\right)  ^{2}$ & if &
$\frac{z_{i-1}+z_{i+1}}{2}\leq t\leq\frac{z_{i}+z_{i+1}}{2},$\\
$0$ & if & $0\leq t\leq\frac{z_{i-1}+z_{i}}{2}$ or $\frac{z_{i}+z_{i+1}}%
{2}\leq t\leq1.$%
\end{tabular}
\right.  \nonumber
\end{align}
Furthermore, by $\pi_i^{a\ast}=0$, we find that those who are not adopted will not gain a positive profit. Thus, without loss of generality, we assume that these prices are $p_{i}^{a\ast}=0$.

First, we assume that the following equation holds: 
\begin{align}\label{prodef}
&p_{i}^{b}>E[\pi_i^{a\ast}]=\int_{\frac{z_{i-1}+z_{i}}{2}}^{\frac{z_{i}+z_{i+1}}{2}}p_{i}^{a\ast}dt\notag\\
&=\int_{\frac{z_{i-1}+z_{i}}{2}}^{\frac{z_{i-1}+z_{i+1}}{2}}%
(t-z_{i-1})^{2}dt+\int_{\frac{z_{i-1}+z_{i+1}}{2}}^{\frac{z_{i}+z_{i+1}}{2}%
}(t-z_{i+1})^{2}dt-\int_{\frac{z_{i-1}+z_{i}}{2}}^{\frac{z_{i}+z_{i+1}}{2}%
}(t-z_{i})^{2}dt.%
\end{align}
This equation implies that $p_{i}^{b}$, an ex-ante price of $i\in M\backslash I$, is higher than the expectation of $p_{i}^{a\ast}$. Note that the right hand side of \eqref{prodef} shows the expected cost for the government.

When the government decides to adopt plan $i$ with $p_i^b$, by this equation \eqref{prodef}, 
we find that its cost is strictly higher than when the government decides not to adopt plan $i$ with $p_i^b$. Thus, it is not appropriate for the government to adopt plan $i$ ex ante. 

Similarly, we find that it is appropriate for the government to adopt plan $i\in M\backslash I$, which suggests an ex-ante price $p_{i}^{b}$ such that  
\begin{equation}
p_{i}^{b}<\int_{\frac{z_{i-1}+z_{i}}{2}}^{\frac{z_{i-1}+z_{i+1}}{2}}
(t-z_{i-1})^{2}dt+\int_{\frac{z_{i-1}+z_{i+1}}{2}}^{\frac{z_{i}+z_{i+1}}{2}
}(t-z_{i+1})^{2}dt-\int_{\frac{z_{i-1}+z_{i}}{2}}^{\frac{z_{i}+z_{i+1}}{2}
}(t-z_{i})^{2}dt. 
\end{equation}
This is because, if they were to add $i$ to $I$ ex ante, they would obtain a higher utility than in this case. 

Next, we show that given the government's best-response behavior, no $i$ has an incentive to suggest $p_{i}^{b}$ lower than his or her ex-post profit: 
\begin{equation}
\int_{\frac{z_{i-1}+z_{i}}{2}}^{\frac{z_{i}+z_{i+1}}{2}}p_{i}^{a\ast}%
dt=\int_{\frac{z_{i-1}+z_{i}}{2}}^{\frac{z_{i-1}+z_{i+1}}{2}}(t-z_{i-1}%
)^{2}dt+\int_{\frac{z_{i-1}+z_{i+1}}{2}}^{\frac{z_{i}+z_{i+1}}{2}}%
(t-z_{i+1})^{2}dt-\int_{\frac{z_{i-1}+z_{i}}{2}}^{\frac{z_{i}+z_{i+1}}{2}%
}(t-z_{i})^{2}dt. %
\end{equation}
If $i$ charges $p_{i}^{b}$ such that 
\begin{equation}\label{notoptimalexante}
p_{i}^{b}<\int_{\frac{z_{i-1}+z_{i}}{2}}^{\frac{z_{i-1}+z_{i+1}}{2}}%
(t-z_{i-1})^{2}dt+\int_{\frac{z_{i-1}+z_{i+1}}{2}}^{\frac{z_{i}+z_{i+1}}{2}%
}(t-z_{i+1})^{2}dt-\int_{\frac{z_{i-1}+z_{i}}{2}}^{\frac{z_{i}+z_{i+1}}{2}%
}(t-z_{i})^{2}dt %
\end{equation}
holds, which is strictly lower than his or her ex-post profit, then the government adopts it ex ante. $i$ obtains a profit equal to $p_{i}^{b}$. However, if $i$ were to deviate to another $p_{i}^{b\prime}$ ex ante such that
\begin{equation}
p_{i}^{b\prime}>\int_{\frac{z_{i-1}+z_{i}}{2}}^{\frac{z_{i-1}+z_{i+1}}{2}%
}(t-z_{i-1})^{2}dt+\int_{\frac{z_{i-1}+z_{i+1}}{2}}^{\frac{z_{i}+z_{i+1}}{2}%
}(t-z_{i+1})^{2}dt-\int_{\frac{z_{i-1}+z_{i}}{2}}^{\frac{z_{i}+z_{i+1}}{2}%
}(t-z_{i})^{2}dt %
\end{equation}
holds, then the government would not adopt it ex ante. Plan $i$'s profit would strictly increase. The researcher obtains
\begin{equation}
\int_{\frac{z_{i-1}+z_{i}}{2}}^{\frac{z_{i-1}+z_{i+1}}{2}}(t-z_{i-1}%
)^{2}dt+\int_{\frac{z_{i-1}+z_{i+1}}{2}}^{\frac{z_{i}+z_{i+1}}{2}}%
(t-z_{i+1})^{2}dt-\int_{\frac{z_{i-1}+z_{i}}{2}}^{\frac{z_{i}+z_{i+1}}{2}%
}(t-z_{i})^{2}dt. %
\end{equation}
Thus, we find that the ex-ante price $p_{i}^{b}$ that \eqref{notoptimalexante} holds is not optimal ex ante.

Finally, we demonstrate that at equilibrium, all the plans indicate $p_{i}^{b}$ such that 
\begin{equation}\label{eqpriceexante}
p_{i}^{b}\geq\int_{\frac{z_{i-1}+z_{i}}{2}}^{\frac{z_{i-1}+z_{i+1}}{2}%
}(t-z_{i-1})^{2}dt+\int_{\frac{z_{i-1}+z_{i+1}}{2}}^{\frac{z_{i}+z_{i+1}}{2}%
}(t-z_{i+1})^{2}dt-\int_{\frac{z_{i-1}+z_{i}}{2}}^{\frac{z_{i}+z_{i+1}}{2}%
}(t-z_{i})^{2}dt %
\end{equation}
holds. Then, the government adopts all the plans or some of them from among those with prices that satisfy this formula~\eqref{eqpriceexante} holding an equal sign.

Furthermore, we find that the condition of the ex-ante price is that an expected net loss is defined as the difference between the realized ideal point and the characteristics of the plans purchased by the government. 

The expected utility of the government is lower when it adopts $i$'s plan that suggests a price such that $p_{i}^{b}$ is strictly higher than the right-hand side of equation~\eqref{eqpriceexante}. Moreover, whether the government adopts plan $i$ that has a price such that $p_{i}^{b}$ equals the right-hand side of equation~\eqref{eqpriceexante}, its expected utility does not change. Thus, we find that this action is the government's best response to research plans described by equation~\eqref{eqpriceexante}.

At the same time, it is not a best response for researcher $i$ to reduce his or her price ex ante in order to be adopted ex ante, because his or her expected profit would be lower. Meanwhile, when the researcher raises the price of his or her plan, the expected profit would not change because the government does not adopt the plan ex ante. Thus, they do not deviate from the price profile given by equation~\eqref{eqpriceexante}. In addition, those whose plans are not adopted do not change their prices ex ante because it does not increase their profit. Thus, they do not deviate from this profile.

We calculate the expected equilibrium price pairs as follows. Note that both ends 1 and N are particular cases because now we consider the [0,1] line.
\begin{align}\label{neqcE}
\begin{cases}
& p^{b\ast}_{1}=\frac{1}{12}[4z_{2}^{3}-(z_{2}-z_{1})^{3}-4z_{1}^{3}],\\
& p^{b\ast}_{i}=\frac{1}{12}[(z_{i+1}-z_{i-1})^{3}-(z_{i+1}-z_{i}%
)^{3}-(z_{i}-z_{i-1})^{3}], \quad 2 \le i \le n-1,\\
&  p^{b\ast}_{n}=\frac{1}{12}[4(1-z_{n-1})^{3}-(z_{n}-z_{n-1})^{3}-4(1-z_{n})^{3}].
\end{cases}
\end{align}
$2 \le i \le n-1$.

\section{Location Game}

In this section, we solve the characteristic choice of research plans. Hereafter, we call this the location game. By equation~\eqref{neqcE}, we obtain $\pi_{i}:(2\leq i\leq n-1)$ as follows.
\begin{align}
\pi_{i} &  =p_{i}=\frac{1}{12}[(z_{i+1}-z_{i-1})^{3}-(z_{i+1}-z_{k}%
)^{3}-(z_{k}-z_{i-1})^{3}] \nonumber\\
= &  \frac{(z_{i-1}-z_{i+1})}{4}\left[  z_{i}-\frac{z_{i+1}+z_{i-1}}%
{2}\right]  ^{2} \nonumber\\
\qquad &  -(\frac{z_{i-1}-z_{i+1}}{16})z_{i+1}^{2}-(\frac{z_{i-1}-z_{i+1}}%
{8})z_{i+1}z_{i-1}-(\frac{z_{i-1}-z_{i+1}}{16})z_{i-1}^{2}+\frac{z_{i-1}%
^{2}z_{i+1}-z_{i-1}z_{i+1}^{2}}{4}%
\end{align}
From the first-order condition for profit maximization, we obtain $z_{i}=\frac{z_{i+1}+z_{i-1}}{2}$. 
Next, we consider plans $1$ and $n$, which are located at both ends. By \eqref{neqcE}, we obtain
\begin{align}
&  \pi_{1}=\frac{1}{12}[4z_{2}^{3}-(z_{2}-z_{1})^{3}-4z_{1}^{3}],\\
&  \pi_{n}=\frac{1}{12}[4(1-z_{n-1})^{3}-(z_{n}-z_{n-1})^{3}-4(1-z_{n})^{3}].
\end{align}

From the first-order conditions for this maximization problem, we obtain
\[%
\begin{cases}
& z_{2}=3z_{1},\\
& \ldots\\
& z_{i+1}-z_{i}=z_{i}-z_{i-1},\\
& \ldots\\
& 3z_{n}=z_{n-1}+2.
\end{cases}
\]
We find that $z_{n}-z_{n-1}=z_{n-1}-z_{n-2}=\cdots=z_{2}-z_{1}=2z_{1}$. 
Thus, we find a sequence of numbers $z_1, z_2, \ldots, z_n$ with common difference $2z_{1}$.
We obtain the general term of this sequence as follows. Now, we find that $z_{i}=2z_{1}i-z_{1},\quad i=1,2,\cdots,n$.
Next, we find that $3z_{n}=z_{n-1}+2 \leftrightarrow 3(2z_{1}n-z_{1})=2(n-1)z_{1}-z_{1}+2
\leftrightarrow z_{1}=\dfrac{1}{2n}$. Thus, we obtain 
\begin{align}
z_{i}=2z_{1}i-z_{1}=\dfrac{2i-1}{2n} \quad i=1,2,\cdots,n.
\end{align}

Substituting $z_{i}=\dfrac{2i-1}{2n}$ for \eqref{neqcE}, we obtain
\begin{align}
&
\begin{cases}
\pi_{1}=p_{1}=\dfrac{1}{12}\left[  4\left(  \dfrac{3}{2n}\right)  ^{3}-\left(
\dfrac{3}{2n}-\dfrac{1}{2n}\right)  ^{3}-4\left(  \dfrac{1}{2n}\right)
^{3}\right],  \notag\\
\\
\pi_{i}=p_{i}=\dfrac{1}{12}\left[  \left(  \dfrac{2i+1}{2n}-\dfrac{2i-3}%
{2n}\right)  ^{3}-\left(  \dfrac{2i+1}{2n}-\dfrac{2i-1}{2n}\right)
^{3}-\left(  \dfrac{2i-1}{2n}-\dfrac{2i-3}{2n}\right)  ^{3}\right],  \notag\\
\\
\pi_{n}=p_{n}=\dfrac{1}{12}\left[  4\left(  1-\dfrac{2n-3}{2n}\right)
^{3}-\left(  \dfrac{2n-1}{2n}-\dfrac{2n-3}{2n}\right)  ^{3}-4\left(
1-\dfrac{2n-1}{2n}\right)  ^{3}\right],  \notag
\end{cases}
\\
&  \Leftrightarrow%
\begin{cases}
\label{loceq}\pi_{1}=\frac{1}{n^{3}},\\
\pi_{i}=\frac{2}{n^{3}},\\
\pi_{n}=\frac{1}{n^{3}}.%
\end{cases}
\end{align}

Till now, we have discussed the necessary condition for a location equilibrium. Now, we check the incentive for a researcher $l\,(1\leq l\leq n)$ to deviate from the location point of the equilibrium defined by \eqref{loceq} to $z_{l}^{\prime}\in\lbrack0,1]\backslash (z_{l-1},z_{l+1})$. 
\[
\pi_{l}^{\prime}=\frac{1}{12}\left[  \left(  \frac{1}{n}\right)  ^{3}-\left(
\dfrac{2i+1}{2n}-z_{l}^{\prime}\right)  ^{3}-\left(  z_{l}^{\prime}%
-\dfrac{2i-1}{2n}\right)  ^{3}\right]  <\frac{1}{12n^{3}}<\pi_{l}%
\]
Next, we check whether any researcher $l$ deviates to a point such that $z_{l}^{\prime}\in\lbrack 0,z_{1})$. 
\[
\pi_{l}^{\prime}=\dfrac{1}{12}\left[  4\left(  \dfrac{1}{2n}\right)
^{3}-\left(  \dfrac{1}{2n}-z_{l}^{\prime}\right)  ^{3}-4z_{l}^{\prime
3}\right]  =\frac{1}{24n^{3}}<\pi_{l}%
\]
Thus, we find that no researcher $l$ deviates.
We obtain the location points of the equilibrium as follows:
\begin{align}
z_{i}^{\ast}=\dfrac{2i-1}{2n}\quad i=1,2,\cdots,n.
\end{align}

\section{Diversity of Research Plans}
In this section, we consider the so-called zero profit condition in our model to obtain product diversity. In the previous section, we assume that fixed cost $F=0$. In other words, under the assumption of $F=0$, every research plan adopted can be implemented and a positive outcome can be attained at equilibrium. However, if some fixed cost $F>0$ is needed before a plan is initiated, then some plans might not be executable because the government cannot afford the fixed cost. For example, it might be necessary for the government to invest in large equipment in advance because the project fund can afford only running costs during the research period. Salop (1979)\cite{salop} calculates the optimal product diversity in the context of spatial competition.

First, we assume that every researcher can gain a reservation value, $0$, when the researcher does not undertake his or her research. In this case, we consider a stage at which everyone deliberates whether he or she will apply for government support. At this stage, when researchers consider that their plans generate a loss because of some positive fixed cost $F$, they stop applying for the contest. We consider a simultaneous move game before researchers choose the locations of their research plans. In this game, there are many potential researchers who have research plans, and they choose whether to enter this competition. If they choose not to enter, they obtain $0$. Entrants $i=1,2,\ldots,n$ play the two-stage game outlined in the previous section. We define each $i$'s profit as follows:
\begin{align*}
&  \pi_{1}=\frac{1}{n^{3}}-F,\\
&  \pi_{i}=\frac{2}{n^{3}}-F,\quad i=2,3,\ldots,n-1,\\
&  \pi_{n}=\frac{1}{n^{3}}-F.
\end{align*}
Based on these profits, each researcher decides whether to submit his or her plans. In a straightforward calculation, we find that it is necessary for the researchers' profits at both the ends, $1,n$, to be non-negative. We define $n^*$ as the number of operating researchers at equilibrium. 

We obtain the following result by solving the zero-profit condition. Note that we assume that a new entrant $n^*+1$ is located at both the ends; that is, its location number is automatically adjusted. At equilibrium, the optimal number of research plans is as follows:
\begin{align}\label{EQ17}
\begin{cases}
n^*=\sqrt[3]{\frac{1}{F}}, \:\text{or}\: \sqrt[3]{\frac{1}{F}}-1,  &\quad\text{if $\left[\sqrt[3]{\frac{1}{F}}\right]=\sqrt[3]{\frac{1}{F}}$},\\
n^*=\left[\sqrt[3]{\frac{1}{F}}\right],  &\quad\text{otherwise}.
\end{cases}
\end{align}
Here, $[x]$ is the largest integer that does not exceed $x$.

\section{Concluding Remarks}
In this study, we consider R$\&$D investment by the government. In our model, there are two subgame perfect equilibria. One is that the government invests in as many plans as it can afford in advance. The other is that the government invests in no plan in advance, and then it invests in only its favorite plan after finding its ideal point. Anyway, it is optimal for the government to invest in all the plans ex ante. More adoption improves the chance of winning a lottery. In other words, our results imply that equally widespread expenditure across all research fields is better than the selection of and concentration in some specific fields. 
At the same time, it is important to cut down on red tape as much as possible. By \eqref{EQ17}, we obtain $dn^{\ast}/dF < 0$. Equation~\eqref{EQ17} shows the number of adopted research plans for the government. This shows that less fixed cost, $F$, is better. This cost includes the cost of office work to make a proposal and to execute a project. This result implies that it is necessary to reduce these administration costs that are required to be paid by the government.

Finally, we discuss the factors of negative outcomes for the current situation in Japan. Recently, economic growth has become the main purpose of Japan's science and technology policy. According to Science and Technology Indicators 2017\cite{mext}, the total sum of R$\&$D expenditure by both the public and private sectors in Japan in 2015 was 18.9 trillion yen (US\$ 170 billion). About 20 years ago, the Japanese government formulated a five-year plan called the Science and Technology Basic Plan ({\it Kagaku Gijyutu Kihon Keikaku}). Currently, the plan has entered its fifth phase, with its latest version issued in 2016. In the current Fifth Basic Plan, science and technology policy is strongly promoted as a major policy for the development of the Japanese economy, which has been suffering from a long recession. The government is seeking innovation that is directly linked to economic growth. Moreover, the government is trying to build processes for dynamic innovation by which R$\&$D findings in one area can be applied to other areas appropriately. The outcome of the government's R $\&$ D investment has two aspects, namely, an outcome useful for the government and a universal achievement for academia. There have been some changes in research expenses in national universities; research funds in natural sciences have increased for both theory and applied development over the past two decades. Despite this trend, with the expansion of R$\&$D by other countries, as per influential papers (Top 10$\%$; Top 1$\%$ papers), Japan declined from the fourth to 10th place in the past decade. According to Benchmarking of Scientific Research 2017 that is included in Science and Technology Indicators 2017, the growth in the number of papers by national universities, which accounts for 50$\%$ of the papers in Japan, has been sluggish since the middle of the 2000s. The promotion of selection and concentration over two decades might have resulted in negative outcomes for research today. 

At last we review the literature on R$\&$D investment in economics. Following the concepts of economics, innovation often creates monopolies with enormous profits, but these gains also provide the incentive for firms to develop new products and processes through R$\&$D. In macroeconomics, relationships between diversity and innovation are the driving force for economic growth; for example, see Aghion and Howitt (1992)\cite{ah1992} and Grossman and Helpman (1994)\cite{gs1994}. Dixit and Stiglitz (1977)\cite{ds1977} provide a microeconomic foundation for the endogenous growth theory. Additional, see Romer (1990)\cite{romer1990} and Grossman and Helpman (1991)\cite{gh1991}. Monopolistic competition plays an important role in the context of innovation and growth. For example, city growth is well captured by this framework. Topics related to cities and agglomerations in urban economics are closely related to economic growth. For example, Behrens et al. (2014)\cite{behrens2014} theoretically and empirically analyzed details on city growth based on the Dixit--Stiglitz model of monopolistic competition. 

On the other hand, how well the monopolistic market works depends on complementarity, that is, flexibility and receptivity. See Matsuyama (1995)\cite{mat1995} for details on the relationship between complementarity and cumulative processes in the Dixit--Stiglitz model of monopolistic competition. Furthermore, R$\&$D investments provide numerous research topics within economics. In the context of R$\&$D, companies themselves are entrepreneurs that earn profits by themselves, which is typical of Dixit and Stiglitz's monopolistic competitive firms. There are several studies considering optimal competitive strategy in this situation, that is, the appropriate timing and investment amount of R$\&$D in competition with other companies. A wide variety of literature in economics relates competition to investment in innovation. Although R$\&$D research is diverse, filed of industrial organization plays a key role in this context. The relationship between the market structure and incentives to invest in R$\&$D has been studied widely in economics. For example, with regard to cost-reducing innovation, Arrow (1962)\cite{arrow1962} finds that a monopolistic firm faces the replacement effect. Dasgupta and Stiglitz (1980)\cite{dasGS1980}, Aghion and Tirole (1994)\cite{at1994}, and Vives (2008)\cite{vives} analyze the incentives for firms to introduce new and improved products. Gilbert and Newbery (1982)\cite{GN1982} point out that preemptive patenting enables firms to protect their market power. As regards the influence of patents on R$\&$D, see, for instance, Katz and Shapiro (1986)\cite{ks1986}. Milgrom and Roberts (1990, 1995)\cite{MR1990,MR1995} provide a theory of complementarity between different activities of the firm, directly applied to the complementarity of firms' innovation activities. With the recent development of structural estimation, research has been shifting from theoretical to empirical studies; see, for example, Igami (2017)\cite{igami}. The microeconomic foundation of economic growth has become more complicated than what has been assumed previously; for example, Dixit and Stiglitz's monopolistic competition. Furthermore, although many economists agree that the relationship between diversity and innovation is the driving force for economic growth, mere recognition would imply the promotion of selection and concentration in some specific fields.

\end{document}